\documentclass{aa} 
\usepackage{graphicx}
\usepackage{color}
\usepackage{graphicx}
\usepackage{lscape}
\usepackage{natbib}
\usepackage[varg]{txfonts}
\usepackage{amssymb}
\usepackage{stmaryrd}
\usepackage{url}
\usepackage{xspace}
\usepackage{color}
\usepackage{siunitx}
\usepackage{textcomp}
\usepackage{amsmath}
\usepackage[usenames,dvipsnames]{xcolor}
\usepackage{longtable}
\usepackage{multicol}
\bibpunct{(}{)}{;}{a}{}{,}
\usepackage[normalem]{ulem}
\usepackage{natbib,twoopt}
\usepackage[breaklinks=true]{hyperref} 
\bibpunct{(}{)}{;}{a}{}{,} 
\makeatletter
\newcommandtwoopt{\citeads}[3][][]{\href{http://adsabs.harvard.edu/abs/#3}%
{\def\hyper@linkstart##1##2{}%
\let\hyper@linkend\@empty\citealp[#1][#2]{#3}}}
\newcommandtwoopt{\citepads}[3][][]{\href{http://adsabs.harvard.edu/abs/#3}%
{\def\hyper@linkstart##1##2{}%
\let\hyper@linkend\@empty\citep[#1][#2]{#3}}}
\newcommandtwoopt{\citetads}[3][][]{\href{http://adsabs.harvard.edu/abs/#3}%
{\def\hyper@linkstart##1##2{}%
\let\hyper@linkend\@empty\citet[#1][#2]{#3}}}
\newcommandtwoopt{\citeyearads}[3][][]%
{\href{http://adsabs.harvard.edu/abs/#3}
{\def\hyper@linkstart##1##2{}%
\let\hyper@linkend\@empty\citeyear[#1][#2]{#3}}}
\makeatother

\def\Gaia{{\it Gaia}}
\newcounter{Rco}
\newcommand{\Ionst}[1]{\setcounter{Rco}{#1}\Roman{Rco}}
\newcommand{\Ion}[2]{\mbox{#1\,{\scriptsize\Ionst{#2}}}}
\newcommand{\Ionw}[3]{\mbox{#1\,{\scriptsize\Ionst{#2}}~$\lambda\,#3$\,\AA}}

\newcommand{\Ionww}[3]{\mbox{#1\,{\scriptsize\Ionst{#2}}~$\lambda\lambda\,#3$\,\AA}}
\newcommand{\pa}{\mbox{Pa\,13}\xspace}

\newcommand{\se}[1]{\mbox{Sect.\,\ref{#1}}}
\newcommand{\logg}{\mbox{$\log g$}\xspace}
\newcommand{\loggw}[1]{\mbox{$\log g\hspace{-0.5mm} =\hspace{-0.5mm} #1$}}

\newcommand{\Teff}{\mbox{$T_\mathrm{eff}$}\xspace}
\newcommand{\Teffw}[1]{\mbox{$\Teff\hspace{-0.5mm} =\hspace{-0.5mm} #1 \,\mathrm{kK}$}}
\newcommand{\ebv}{$E_\mathrm{B-V}$\xspace}

\newcommand{\Lsol}{$\mathrm{L}_\odot$\xspace}
\newcommand{\Msol}{$\mathrm{M}_\odot$\xspace}
\newcommand{\Rsol}{$\mathrm{R}_\odot$\xspace}

\begin{document} 

\title{Two hot pre-white dwarfs inside the red-giant-branch planetary nebula Pa~13}
\subtitle{Double core evolution or common envelope-induced rejuvenation?}
\titlerunning{Two hot pre-white dwarfs inside Pa~13}

\author{Nicole Reindl \inst{1}
  \and David Jones\inst{2, 3}
  \and Todd Hillwig\inst{4}
  \and Marcelo M. Miller Bertolami\inst{5}
  \and Matti Dorsch\inst{6}
  \and Nicholas Chornay\inst{7}
  \and Max Pritzkuleit\inst{6}
  }
  
\offprints{Nicole\,Reindl\\ \email{nreindl885@gmail.com}}

\institute{Zentrum für Astronomie der Universität Heidelberg, Landessternwarte, Königstuhl 12, D-69117 Heidelberg, Germany
\and Instituto de Astrof\'{i}sica de Canarias, E-38205 La Laguna, Tenerife, Spain
\and Departamento de Astrof\'{i}sica, Universidad de La Laguna, E-38206 La Laguna, Tenerife, Spain
\and Department of Physics and Astronomy, Valparaiso University, Valparaiso, IN 46383, USA
\and Instituto de Astrof\'{i}sica de La Plata, UNLP-CONICET, La Plata, 1900 Buenos Aires, Argentina
\and Institut f\"ur Physik und Astronomie, Universit\"at Potsdam, Haus 28, Karl-Liebknecht-Str. 24/25, D-14476 Potsdam-Golm, Germany
\and Department of Astronomy, University of Geneva, Chemin d’Ecogia 16, 1290 Versoix, Switzerland
}

\date{Received 19 January 2026 \ Accepted 16 March 2026}

\abstract
{Close binary central stars of planetary nebulae (PNe) offer a unique window for investigating the conditions immediately following the ejection of a common envelope (CE). Double eclipsing and double-lined double systems are particularly valuable as they provide minimally model-dependent constraints on fundamental binary parameters. In this context, we report that the nucleus of Pa~13 ($P=0.3988$\,d) belongs to this rare class of systems and present a comprehensive analysis of its double-degenerate binary. We performed a two-component non-local thermodynamic equilibrium spectral analysis based on phase-resolved X-Shooter spectroscopy, multi-band light-curve modeling, spectral energy distribution fitting, as well as a kinematic analysis. Both stars are found to be hot pre-white dwarfs, with Star~1 being cooler but larger (\Teffw{50.0}, $R=0.40$\Rsol) than Star~2 (\Teffw{75.0}, $R=0.16$\Rsol). The weakness of spectral lines of Star~2 made both the atmospheric and radial velocity (RV) analyses challenging, and we uncovered a strong sensitivity of the assumed surface ratio to its derived RV curve. Yet, the RV curve and Kiel mass of Star~1 ($M_1=0.41\pm0.02$\,\Msol) could be determined precisely, allowing for a dynamical mass determination of Star~2 ($M_2=0.39\pm0.04$\,\Msol). 
Moreover, we uncovered that Pa~13 exhibits a small but significant orbital eccentricity ($e = 0.02\pm0.01$), making it only the second post-CE binary central star with a measured eccentricity. Our kinematic analysis shows that Pa~13 belongs to the Galactic halo, implying a system age of $\approx 11$\,Gyrs. We conclude that Pa~13 provides hitherto the strongest evidence that PNe can be observed around post-red giant branch stars. Immediately after the CE-ejection, Star~1 likely still filled its Roche lobe, suggesting that Pa~13 is a more evolved, detached descendant of over-contact double-degenerate systems such as \object{Hen~2-428}. Since the mass ratio of Pa~13 is close to unity the system may have formed through double-core CE evolution. Alternatively, there must exist an efficient CE-induced rejuvenation mechanism capable of reheating the cool white dwarf in the binary, as already indicated by the Hen~2-428 system.}


\maketitle

\section{Introduction}
\label{sec:intro}
The study of close double-degenerate binaries is of considerable importance as these systems do serve as powerful sources of gravitational waves \citep{Evans+1987, Lipunov+1987, Hils+1990, Nelemans+2001c, Burdge+2019, Li+2020}, and contribute to our understanding of the progenitors of Type Ia supernovae \citep{Webbink1984} and to the origin of hyper-velocity stars \citep{Neunteufel+2019, Shen+2025}. Moreover, double-degenerates constitute essential laboratories for investigating close binary evolution, including the common envelope (CE) phase \citep{Nelemans+2005, Nelemans+2025}. Beyond their role as potential progenitors of thermonuclear supernovae, double-degenerate mergers may also give rise to ultra-massive white dwarfs (WDs, \citealt{Wu+2022}), highly magnetic WDs \cite{Garcia-Berro+2012}, and a variety of other exotic stars \citep{Zhang+2012a, Zhang+2012b, WernerRauch2015, Kawka+2020, Dorsch+2022, Mackensen+2025}. 

Close binary central stars (CSs) of planetary nebulae (PNe) provide one of the best means of observationally exploring CE evolution and assessing current CE models. This is because the ejected envelope is still visible (the PN) and the post-CE binary can be examined prior to the onset of significant subsequent evolutionary processes, such as mass transfer or magnetic and gravitational braking. For instance, several post-CE systems have been found in which the PN symmetry axis lies perpendicular to the binary orbital plane \citep{Hillwig+2016b, Munday+2020}, consistent with theoretical predictions (e.g. \citealt{Garcia-Segura+2018}). Furthermore, it was uncovered that in post-CE systems containing a main-sequence star, the main-sequence companion is almost always hotter and larger than expected for an isolated main-sequence star of the same mass \citep{Jones+2015, Jones+2020, Jones+2022}. The observed heating and/or inflation is generally attributed to the companion’s incomplete thermal readjustment following the brief but intense episode of accretion immediately preceding the CE phase, yet the extreme irradiation from the hot CS may also play a role \citep{DeMarco+2008, Jones+2022}. This inflation is not observed in older post-CE systems in which the PN has already dissipated and the primary has entered the WD cooling sequence \citep{Parsons+2010}. Consequently, post-CE CSPNe constitute particularly valuable systems for investigating the immediate response of a companion to the CE phase. 
Another notable result from detailed analyses of post-CE CSPNe hosting irradiated low-mass main-sequence companions is the emergence of observational evidence indicating that some PNe originate from a CE phase that occurred on the red-giant branch (RGB; \citealt{Afsar+Ibanoglu2008, Hillwig+2016, Hillwig+2017, Jones+2020, Jones+2022}). This finding challenges the classical paradigm, which predicts that only post–asymptotic giant branch (AGB) stars can become CSPNe.

Currently, about 100 strong candidates of post-CE CSPNe have been detected via photometric variability\footnote{\url{https://www.drdjones.net/bcspn/}}. The majority ($\approx 66$\%) of them being irradiation effect systems containing a hot CS and an irradiated low-mass main-sequence star. Such systems are easily detected in photometric surveys due to the large-amplitude variations arising from the high temperature contrast between the irradiated and non-irradiated hemispheres of the low-mass companion. In about 30\% of the photometrically variable post-CE CSPNe binary candidates ellipsoidal variability is observed, which is thought to be caused by the tidal deformation of the hot CSs by a close compact companion. Additional evidence for close, double-degenerate CSPNe systems comes from a handful of single-lined binaries \citep{Rodriguez-Gil+2010, Boffin+2012, Miszalski+2018b, Miszalski+2019a, Miszalski+2019b, Miszalski+2019c}.

To make solid statements about the current, past and future conditions of the binary, the system parameters need to be known accurately. In most cases the atmospheric parameters of the hot CSs are derived from optical spectroscopy, but in the case of very hot CSs (\Teff$\gtrapprox 70$\,kK) these can often involve large systematic uncertainties (e.g. \citealt{Reindl+2014b, Werner+2018a}). This often entails a mismatch of the spectroscopic and \Gaia\ distances or equivalently a mismatch of the Kiel and gravity masses\footnote{The gravity mass is defined as $M_{\mathrm{g}}=\frac{g_{\mathrm{spec}} R_{\mathrm{Gaia}}^2}{G}$, where $g_{\mathrm{spec}}$ is the surface gravity derived from spectroscopy, $R_{\mathrm{Gaia}}$ the stellar radius obtained by spectral energy distribution fitting assuming the distance from the \Gaia\ mission, and $G$ is the gravitation constant.} \citep{Napiwotzki+2001, Schoenberner+2018, SchoenbernerSteffen2019, Frew+2016, Reindl+2023, Reindl+2024}. Conclusions regarding the nature of the companion are typically drawn from the interpretation and modeling of the system’s light curve, ideally in combination with the radial velocity (RV) curves of both stars. In this regard, double-lined and double eclipsing systems are most valuable since the period and inclination angle can be derived from light curve modeling, and phase-resolved spectroscopy can provide the RV semi-amplitudes, $K_1$ and $K_2$, of both stars. With those values, the dynamical masses of both stars can be determined, which -- together with the radii from the light curve fitting -- provide valuable and independent tests for stellar evolutionary calculations and the mass-radius relationship of WDs \citep{Parsons+2017, Higl+Weiss2017}.

However, since both stars are required to have comparable luminosities and must be observed at a high inclination angle, the chances of detecting a double eclipsing, double-lined and double-degenerate system are considered very low \citep{Rebassa-Mansergas+2019}  

Considering that there are currently 3962 Galacitc PN candidates listed in the HASH database, this translates to a relative fraction of 0.05\%. In comparison, among the about 359\,000 WD \citep{GentileFusillo+2021} and 61\,585 hot subdwarf candidates \citep{Culpan+2022} only 13 and one double eclipsing and double-lined systems are know, respectively. This in turn implies current detection rates of 0.003\% and 0.002\% for WDs and hot subdwarfs, respectively. Thus, the detection rate among CSPNe is enhanced by a factor of $17-25$ compared to (pre-)WDs that lack a PNe. Consequently, such systems are rare, with only 13 double-lined, double-eclipsing double WDs known to date \citep{Munday+2024, AntunesAmaral+2024}. Among pre-WDs, the only known double-eclipsing, double-lined and double-degenerate system are the double hot subdwarf binary \object{SDSS\,J135713.14-065913.7} \citep{Finch+2019} and the nucleus of the PN Hen~2-428. The latter is an over-contact system, which means that both stars still share a CE, and makes the spectral analysis of the system challenging \citep{SG+2015, Reindl+2020}. 

Here we announce the discovery of the double lined nature of the double-degenerate and double eclipsing nucleus of the PN Pa~13 (\object{PNG\,041.4$-$09.6}). The PN was discovered by \cite{Kronberger+2014} and is listed as a true PN in the HASH data base with a diameter of 29\,arcsec.
The centrally located nucleus itself was identified in the second \Gaia\ data release as \object{\Gaia\ DR2 4289787771600957824} by \citet{Chornay2020}.
Utilizing archived photometric data from the Zwicky Transient Facility (ZTF) survey, \cite{Chen+2025} found a consistent period of 0.3988\,d in the g- and r-band light curves of the CS. They reported that the light curve is dominated by two eclipses with different minima, and that there seems to be some ellipsoidal modulation present. The double eclipsing nature of the ZTF light curves of Pa\,13 was also found by \cite{Bhattacharjee+2025}, who reported the same period. 


\section{Observations} 
\label{sec:obs}

\subsection{Photometry} 
\label{sec:photo}

We obtained ZTF light curves (ZTF, \citealt{Bellm+2019, Masci+2019}) from the data release 23 for Pa~13. They provide 413 and 1013 data points in the g- and r-band respectively. 

Further g-band photometry, targeting primary and secondary eclipses, was obtained on the nights of 12--14 July 2021 using the Wide Field Camera (WFC) instrument mounted on the 2.5-m Isaac Newton Telescope (Programme ID: 96-INT19/21A). In total, 230 exposures of 45s were obtained and the data were reduced using standard astropy routines before performing differential aperture photometry against non-variable field stars. The data were then normalised to the same apparent magnitude scale as the ZTF data.

Similarly, a total of 70 i-band exposures of 300s were obtained using the 1-m SARA-La Palma Telescope (formerly the Jacobus Kapteyn Telescope, \citealt{Keel+2017}) on the nights of 28 May and 11 August 2021. These data were reduced using standard IRAF routines with differential photometry calculated against non-variable field stars.

Combining the data, we derive an ephemeris of:
\begin{equation}
\mathrm{HJD}_\mathrm{min}=2460610.3864(4) + 0.3987674(8) E
\end{equation} 
where HJD$_\mathrm{min}$ is the Heliocentric Julian Date of the mid-point of the deeper eclipse and $E$ is an integer representing the number of complete orbits since the first time of minimum.

\subsection{Spectroscopy} 
\label{sec:spectra}

\begin{table}[t]
\caption{X-Shooter observations of Pa\,13.}
\label{tab:obs}
\centering
\begin{tabular}{l l l l r} 
\hline\hline
\noalign{\smallskip}
 HJD$_\mathrm{middle}$ & $t_\mathrm{exp}$ & SNR & Phase & $RV_1$\\
            & [s] &   &    & [km/s]\\
\noalign{\smallskip}
\hline 
\noalign{\smallskip}
2459768.626406	&	1867	&	9	&	0.10 & $106.4\pm3.0	$\\
2459768.648958	&	1867	&	13	&	0.15 & $126.8\pm3.3	$\\
2459768.671508	&	1867	&	16	&	0.21 & $151.0\pm2.8$\\
2459768.695427	&	1867	&	20	&	0.27	& $	156.9\pm2.3	$\\
2459768.717957	&	1867	&	26	&	0.33	& $	152.0\pm2.4	$\\
2459768.740502	&	1867	&	36	&	0.38	& $	121.6\pm2.3	$\\
2459768.764346	&	1867	&	23	&	0.44	& $	100.1\pm3.3	$\\
2459768.871251	&	1729	&	8	&	0.71	& $	-89.4\pm5.0	$\\
\noalign{\smallskip}
\hline 
\noalign{\smallskip}
2460134.568271	&	1867	&	51	&	0.78	& $	-99.1\pm1.8	$\\
2460134.590814	&	1867	&	51	&	0.84	& $	-83.0\pm1.7	$\\
2460134.613359	&	1867	&	54	&	0.89	& $	-47.8\pm	1.7	$\\
2460134.637253	&	1867	&	53	&	0.95	& $	-4.1\pm2.4	$\\
2460134.659794	&	1867	&	54	&	0.01	& $	58.6\pm2.3	$\\
2460134.682338	&	1867	&	56	&	0.06	& $	93.4\pm1.8	$\\
2460134.706176	&	1867	&	56	&	0.12	& $	128.8\pm1.4	$\\
2460134.728711	&	1867	&	55	&	0.18	& $	149.8\pm1.5	$\\
2460134.751257	&	1867	&	52	&	0.24	& $	158.9\pm1.4	$\\
2460134.774893	&	1867	&	57	&	0.30	& $	145.4\pm1.5	$\\
2460134.797450	&	1867	&	53	&	0.35	& $	132.1\pm1.6	$\\
2460134.821246	&	1867	&	41	&	0.41	& $	100.5\pm2.3	$\\
2460134.847034	&	1867	&	38	&	0.48	& $	63.5\pm3.1	$\\
2460134.869571	&	1867	&	36	&	0.53	& $	-5.8\pm3.1	$\\
2460134.892128	&	1867	&	23	&	0.59	& $	-37.6\pm4.1	$\\
\noalign{\smallskip}
\hline
\end{tabular}
\tablefoot{For each spectrum the Heliocentric Julian Day at middle of the exposure, exposure time, and SNR between 5100$-$5300\AA\ is given. The quoted uncertainties are only from the template cross-correlation and do not include any uncertainty associated with the wavelength calibration.
}
\end{table}

After we independently discovered the double eclipsing nature of the nucleus of \pa, we requested classification spectroscopy in order to check if the system is potentially also double lined. Two epochs of spectroscopy were obtained under Director's Discretionary Time (ProgID: GTC05-21ADDT, PI: Jones) with the 10.4~m Gran Telescopio Canarias (GTC) using the Optical System for Imaging and low Resolution Integrated Spectroscopy (OSIRIS). The spectra were obtained with with R2000B grating and a 0.6" slit resulting in a resolving power of $R\approx2165$ over the wavelength range 3950 $\lesssim \lambda \lesssim$ 5700 \AA. The spectra were obtained on the nights 10 July 2021 and 10 August 2021, with three consecutive spectra each with an exposure time of 900\,s taken on both nights. The data were reduced using the PypeIt \textsc{python} package \citep{pypeit}.\\
In addition, we obtained another spectrum close to the expected maximum RV separation at the twin 8.4~m Large Binocular Telescope (LBT) using the Multi-Object Double Spectrographs (MODS, \citealt{Pogge+2010}) on June 9, 2021 using an exposure time of $2\times 900$\,s (ProgID: RDS-2021B-004, PI: Reindl). MODS provides two-channel grating spectroscopy by using a dichroic that splits the light at $\approx5650$\,\AA\ into separately optimized red and blue channels. The spectra cover the wavelength region $3330 - 5800$\,\AA\ and $5500 - 10\,000$\,\AA\ with a resolving power of $R\approx 1850$ and $2300$, respectively. We reduced the spectra using the modsccdred\footnote{\url{https://github.com/rwpogge/modsCCDRed}} \textsc{PYTHON} package \citep{Pogge2019} for basic 2d CCD reductions, and the modsidl\footnote{\url{https://github.com/rwpogge/modsIDL}} pipeline \citep{Croxall+2019} to extract 1d spectra and apply wavelength and flux calibrations. Based on these GTC/OSIRIS and LBT/MODS spectra, we found that the nucleus of \pa is indeed double lined as evident from \Ionw{He}{2}{4686}.

Subsequently, we requested phase-resolved spectroscopy with the X-shooter instrument \citep{Vernet+2011} at ESO’s Very Large Telescope (VLT). The spectra cover the wavelength range $3000-10\,000\,\AA$ and have a resolving power of $R\approx 10\,000$.
The first attempt was carried out on July 8, 2022 (ProgID: 0109.D-0237(A); PI: Reindl). These data were reduced using the ESO Reflex pipeline. However, due to poor weather conditions during that night, only eight spectra could be obtained and with a relatively poor signal-to-noise ratio (SNR). On July 9, 2023 a second and successful attempt was made to obtain phase-resolved spectroscopy X-shooter (ProgID: 0111.D-2143(A); PI: Reindl). We downloaded the extracted, wavelength- and flux-calibrated 1D spectra from the ESO science archive.\\
The X-shooter UVB arm spectra display photospheric lines of \Ion{H}{1} and \Ion{He}{2}, as well as a weak \Ionww{He}{1}{4471} line. H\,$\beta, \gamma, \delta$, and \Ionw{He}{2}{4686} are blended by residual nebular emission lines. 
In addition nebular lines of [\Ion{Ne}{3}] $\lambda 3869\,\AA$ and [\Ion{O}{3}] $\lambda\lambda 4363, 4959, 5007\,\AA\AA$, a weak diffuse interstellar band (DIB) at $\lambda 4430\,\AA$, along with the interstellar Ca H \& K absorption are visible in the UVB arm (see also Fig.~\ref{fig:ubv}). In Table~\ref{tab:obs} we provide an overview of the UVB spectra and list for each observation the HJD at middle of the exposure, the exposure time, and SNR as measured between 5100$-$5300\AA. The VIS arm spectra only display the interstellar Na I doublet absorption lines and a weak \Ionw{He}{1}{5876} line, which is, however, blended with a nebular line. Depending on the SNR of the observation, the absorption wings of photospheric H\,$\alpha$ are apparent, but in most cases the line is dominated by nebular emission. Consequently, we restricted ourselves to the UVB arm observation for the spectral analysis. We also exclude the MODS and OSIRIS data for the spectral analysis to avoid zero-point offsets between the different instruments. Furthermore, the relatively low spectral resolution of the OSIRIS and MODS data, combined with the weakness of the spectral lines of Star~2, makes them unsuitable for precise RV measurements.

\section{Spectral analysis}
\label{sec:spec}

\subsection{System velocity}
\label{sec:v_sys}
To evaluate possible residual zeropoint shifts within the individual X-shooter exposures and runs, we determined the system velocity for each exposure. This was done by deriving the RVs from [\Ion{Ne}{3}] $\lambda 3869\,\AA$ and [\Ion{O}{3}] $\lambda 4363\,\AA$, which are some of the few residual nebular lines located near the center of the UVB arm, which was also used to determine the photospheric RVs. In addition these lines are narrow and not blended with photospheric absorption lines. The [\Ion{O}{3}] $\lambda\lambda 4959, 5007\,\AA\AA$ lines appear double-peaked -- likely as a result of the background subtraction -- and were not employed for the determination of the system velocity. 
The system velocity was then derived by by fitting a Voigt profile to the observed nebular emission line in each spectrum. By that we find $v_{\mathrm{sys}}=33.7\pm0.8$\,km/s. We also note, the system velocity derived from the spectra obtained in 2022 ($v_{\mathrm{sys}}=34.1\pm1.7$\,km/s) agrees with the one derived from the 2023 spectra ($v_{\mathrm{sys}}=33.5\pm1.8$\,km/s). We therefore conclude that wavelength calibration uncertainties do not dominate the RV error budget.

\begin{figure}[t]
 \centering 
\resizebox{\hsize}{!}{\includegraphics{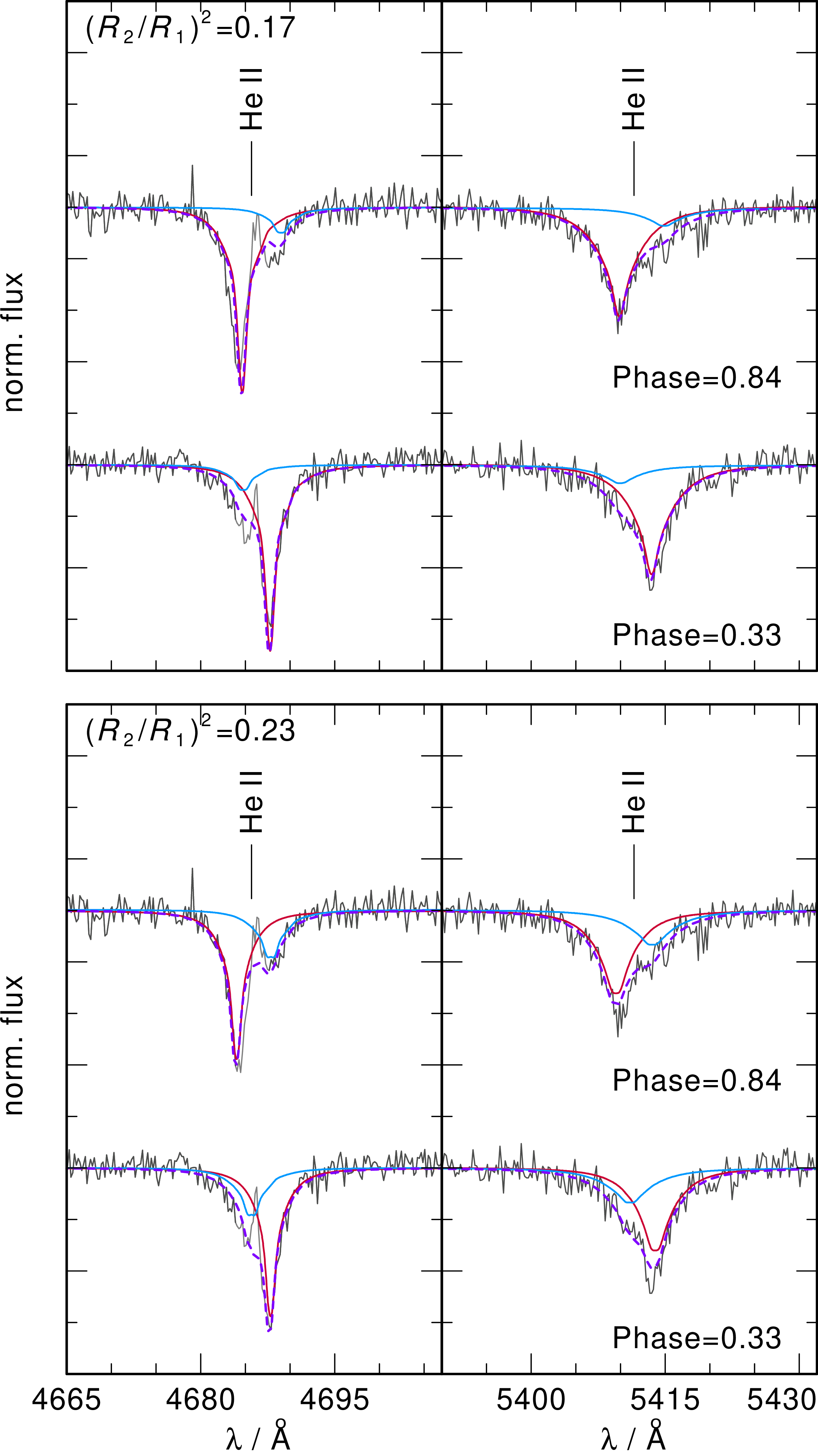}}
 \caption{Two examples X-Shooter spectra (gray) taken close to maximum RV separation, respectively. Light gray regions indicate the location of a weak residual nebular line, which have been excluded from the fit. The red and blue lines represent the contribution of Star~1 and Star~2, respectively, and the dashed, purple line shows the combined fit combined fit. The upper panel show the predicted line profiles assuming a surface ratio of 0.17, while the lower panel the line profiles assuming a surface ratio of 0.23. Applying a consistent approach to all spectra, the resulting derived RV semi-amplitudes of Star~2 are $K_2=143$\,km/s and $K_2=78$\,km/s for an assumed surface ratio of 0.17 and 0.23, respectively.}
\label{fig:RVs}
\end{figure}

\subsection{Atmospheric parameters and radial velocities}
\label{sec:atmo}

We employed the metal-free non-local thermodynamic equilibrium (NLTE) model grid calculated by \cite{Reindl+2016} and performed two-component $\chi^2$ spectral fits to the X-Shooter UVB spectra. Several absorption lines of H and He were considered in the fit, but we removed the line cores of \Ionw{He}{2}{4686} and the Balmer lines (i.e. $\pm2$\,\AA) from the line center, which show very weak residual nebular emission. 
Since the spectral fits are not sensitive to the surface ratio, we kept the latter fixed at a value predicted by the light curve fitting. The initial light curve fit predicted a surface ratio of $(R_2/R_1)^2=0.1$, however, this value increased as we included the results from the spectroscopic analysis (atmospheric parameters and RVs) into the light curve fitting. Therefore, the atmospheric parameters and RVs were determined iteratively by taking into account the results from the light curve fitting. In the fit we then derived the effective temperatures, surface gravities, and He abundances from all X-shooter simultaneously. The RVs were allowed to vary freely within the individual exposures. We note that we kept the surface ratio fixed also for exposures affected by the eclipse, since the eclipses are very shallow ($\approx 0.2$\,mag) and the derived atmospheric parameter of both stars were found to be insensitive to such minor changes in the surface ratio.

For Star~1 -- which has the stronger lines -- the atmospheric parameters as well as the RVs could be constrained precisely and did not depend strongly on the range of surface ratios ($(R_2/R_1)^2=0.1-0.23$) that were considered. We derive \Teff$=50.0_{-0.2}^{+1.5}$\,kK, \loggw{4.85\pm0.07}, and $\log({\mathrm{He/H})}=-1.14\pm0.06$, which is about solar. The formal fitting errors are small, thus we determined the errors by considering the results from the various surface ratios that were assumed.

Yet, we found that the atmospheric parameters of Star~2 -- which has the weaker lines -- are very sensitive to the start parameters of the fit, as well as the assumed light curve ratio. Fixing \logg to the value predicted by the light curve models resulted in a \Teff for Star~2 similar to that of Star~1. However, since the difference in the depths of the eclipses clearly indicates distinct effective temperatures for both stars, we fixed the effective temperature of Star~2 to the value predicted by the light curve model and only allowed \logg and the He abundance to vary freely. The derived values for the He abundance stayed close to the solar value ($\log({\mathrm{He/H})}=-1.00\pm0.30$), but the spectroscopically determined value for \logg always remained $0.3-0.5$~dex below the value derived from the light curve fit (\loggw{5.5}). We speculate that this problem might be related to the Balmer line problem, which occurs for stars whose \Teff exceeds $\approx 70$\,kK and describes the failure to achieve a consistent fit to all Balmer (and \Ion{He}{2}) lines simultaneously \citep{Werner1996}. In some cases this problem can be overcome by the inclusion of metal opacities in the model atmosphere calculations, and several studies using such models have shown that the resulting values for \Teff and \logg can be significantly different from an analysis that uses metal-free models (e.g., \citealt{Filiz+2024}).

For Star~1 the systematic error based on the surface ratio on the derived RV semi-amplitude, $K_1$, is only a few km/s, i.e. $K_1$ varies between 124 and 135\,km/s if a circular orbit is assumed. However, we found a drastic effect of the assumed surface ratio on the derived RVs of Star~2 and consequently on $K_2$. Assuming a circular orbit, we derive $K_2=163$\,km/s for a surface ratio of $(R_2/R_1)^2=0.1$, which implies $M_1 > M_2$, for $(R_2/R_1)^2=0.17$, we derive $K_2=143$\,km/s (implying $M_1 \approx M_2$), while for for $(R_2/R_1)^2=0.23$ $K_2$ drops down to 78\,km/s (implying $M_1 < M_2$). This dependency becomes understandable when looking at Fig.~\ref{fig:RVs}, where the best-fitting line profiles for the \Ionww{He}{2}{4686, 5412} lines that result from surface ratios of 0.17 (upper panel) and 0.23 (lower panel) are shown for two different phases (0.84 and 0.33). For an increasing surface ratio, the lines of Star~2 (shown in blue) become stronger, and consequently the RV shift that is needed to reproduce the observation becomes lower. Thus, $K_2$ naturally decreases with increasing surface ratio.

We conclude that neither the atmospheric parameters, nor the RV curves of Star~2 can be determined reliably based on the current spectroscopic observations. Therefore, we adopt the parameters obtained from the light curve analysis for Star~2. For the strong-lined Star~1, we adopt the spectroscopically determined atmospheric parameters and RV curve. We also point out that Star~1 does not display the Balmer line problem as can be seen in Fig.~\ref{fig:ubv}, where we show the UVB X-Shooter spectrum taken at phase 0.84 is compared to our best-fitting TMAP models. The atmospheric parameters of both stars including their errors are summarized in Table~\ref{tab:para} and the RVs as determined assuming a surface ratio of 0.17 are listed in Table~\ref{tab:obs}.

\subsection{Kiel mass}
\label{sec:kiel}

\begin{figure}[t]
\centering 
\resizebox{\hsize}{!}{\includegraphics{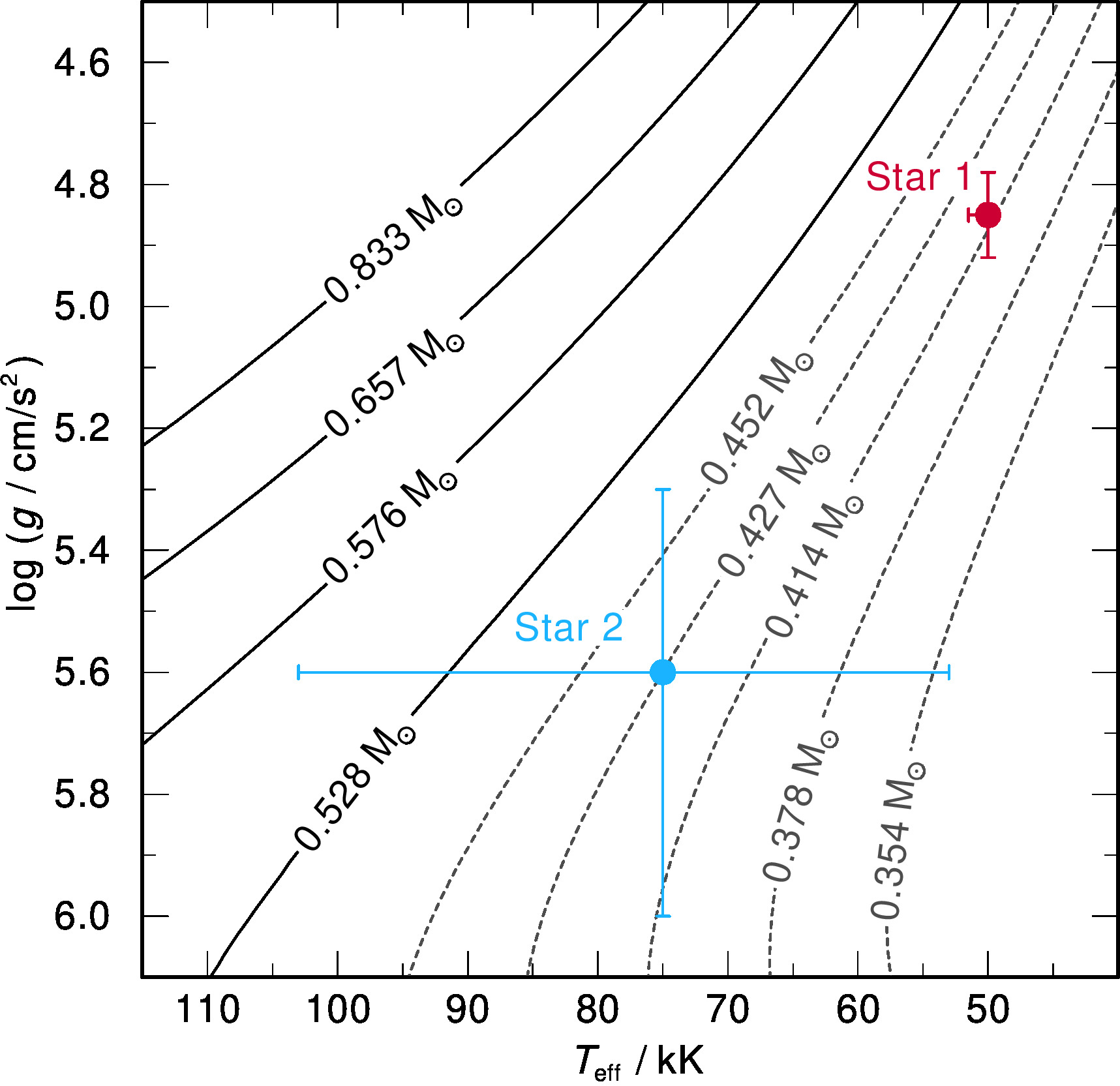}}
\caption{Kiel diagram, showing the position of the two CSs of Pa~13 compared to H-shell burning post-AGB tracks (black lines) from \cite{MillerBertolami2016} and post-RGB tracks (dashed, gray lines) from \cite{Hall+2013}.}
\label{fig:kiel}
\end{figure}

In Fig.~\ref{fig:kiel} we show the position of the two binary CSs of Pa~13 in a Kiel diagram compared to H-shell burning post-AGB tracks (black lines) from \cite{MillerBertolami2016} and post-RGB tracks (dashed, gray lines) from \cite{Hall+2013}. Star~1 clearly lies in the post-RGB region. Using the post-RGB evolutionary sequences from \citep{Hall+2013} and the {\tt griddata} (linear) interpolation function in \textsc{SciPy} \citep{2020SciPy-NMeth}, we derived a Kiel mass of $0.41\pm0.02$\,\Msol for Star~1. 

\section{Light and RV curve fitting} 
\label{sec:lc}

Simultaneous light and RV curve modeling of the CSs of Pa~13 was carried out using \textsc{phoebe2} \citep{2016ApJS..227...29P,2020ApJS..247...63J,2020ApJS..250...34C}. The spectrum and limb-darkening of both stars were approximated using the same atmosphere models used in \se{sec:atmo} as implemented in \textsc{phoebe2} \citep{jones2022,jones2026}, with the effective temperature, surface gravity and He abundance of Star~1 fixed to the values derived there. For Star~2, we also fixed the He abundance to the value determined previously, but its \Teff and \logg were allowed to vary freely over a wide range of values.

\begin{table}[t]
\caption{Orbital and stellar parameters of Pa~13.}
\begin{tabular}{l r@{\,$\pm$\,} l r@{\,$\pm$\,} l} 
\hline\hline
\noalign{\smallskip}
         & \multicolumn{2}{c}{Star~1}  & \multicolumn{2}{c}{Star~2}   \\
\noalign{\smallskip}
\hline
\noalign{\smallskip}
$P$ [days]$^{(a)}$    & \multicolumn{4}{c}{0.3987674\,$\pm$\,0.0000008}\\
\noalign{\smallskip}
$\gamma$ [km/s]    & \multicolumn{4}{c}{33.7\,$\pm$\,0.8}\\
\noalign{\smallskip}
$q \equiv M_2/M_1$   & \multicolumn{4}{c}{$0.95^{+0.15}_{-0.14}$}\\
\noalign{\smallskip}
$i $ [\textdegree]   & \multicolumn{4}{c}{$81.2^{+0.6}_{-3.6}$}\\
\noalign{\smallskip}
$e $   & \multicolumn{4}{c}{$0.02\pm0.01$}\\
\noalign{\smallskip}
$a $ [\Rsol]      & \multicolumn{4}{c}{2.1$\pm$0.1}\\
\noalign{\smallskip}
$d$ ({\it Gaia}) [kpc] & \multicolumn{4}{c}{$6.9^{+4.8}_{-2.1}$}\\
\noalign{\smallskip}
$E(44-55)$ [mag]    & \multicolumn{4}{c}{$0.405 \pm 0.005$} \\
\noalign{\smallskip}
$T_{\rm eff}$ [K]  & \multicolumn{2}{c}{$50.0_{-0.2}^{+1.5}$} & \multicolumn{2}{c}{$75.0_{-22.0}^{+28.0}$} \\
\noalign{\smallskip}
$\log{g}$         & $4.85$&$0.07$ & \multicolumn{2}{c}{$5.60^{+0.30}_{-0.40}$} \\
\noalign{\smallskip}
$\log({\mathrm{He/H})}$    & $-1.14$&$0.06$ & $-1.00$&$0.30$\\
\noalign{\smallskip}
$M$ [\Msol]   &     0.41&0.02 & 0.39&0.04\\
\noalign{\smallskip}
$L$ [\Lsol]      & \multicolumn{2}{c}{$903_{-183}^{+271}$} & \multicolumn{2}{c}{$731_{-645}^{+5120}$}\\
\noalign{\smallskip}
$R$ [\Rsol]   & 0.40&0.04 & \multicolumn{2}{c}{$0.16_{-0.07}^{+0.08}$} \\
\noalign{\smallskip}
$R_{\mathrm{Gaia}}$ [\Rsol]  & \multicolumn{2}{c}{$0.45^{+0.32}_{-0.14}$} & \multicolumn{2}{c}{$0.19^{+0.13}_{-0.06}$} \\
\noalign{\smallskip}
\hline
\end{tabular}
\tablefoot{ He abundances are given in logarithmic number fractions relatively to H.
}
\label{tab:para}
\end{table}

\begin{figure*}[t]
\sidecaption
\includegraphics[width=12cm]{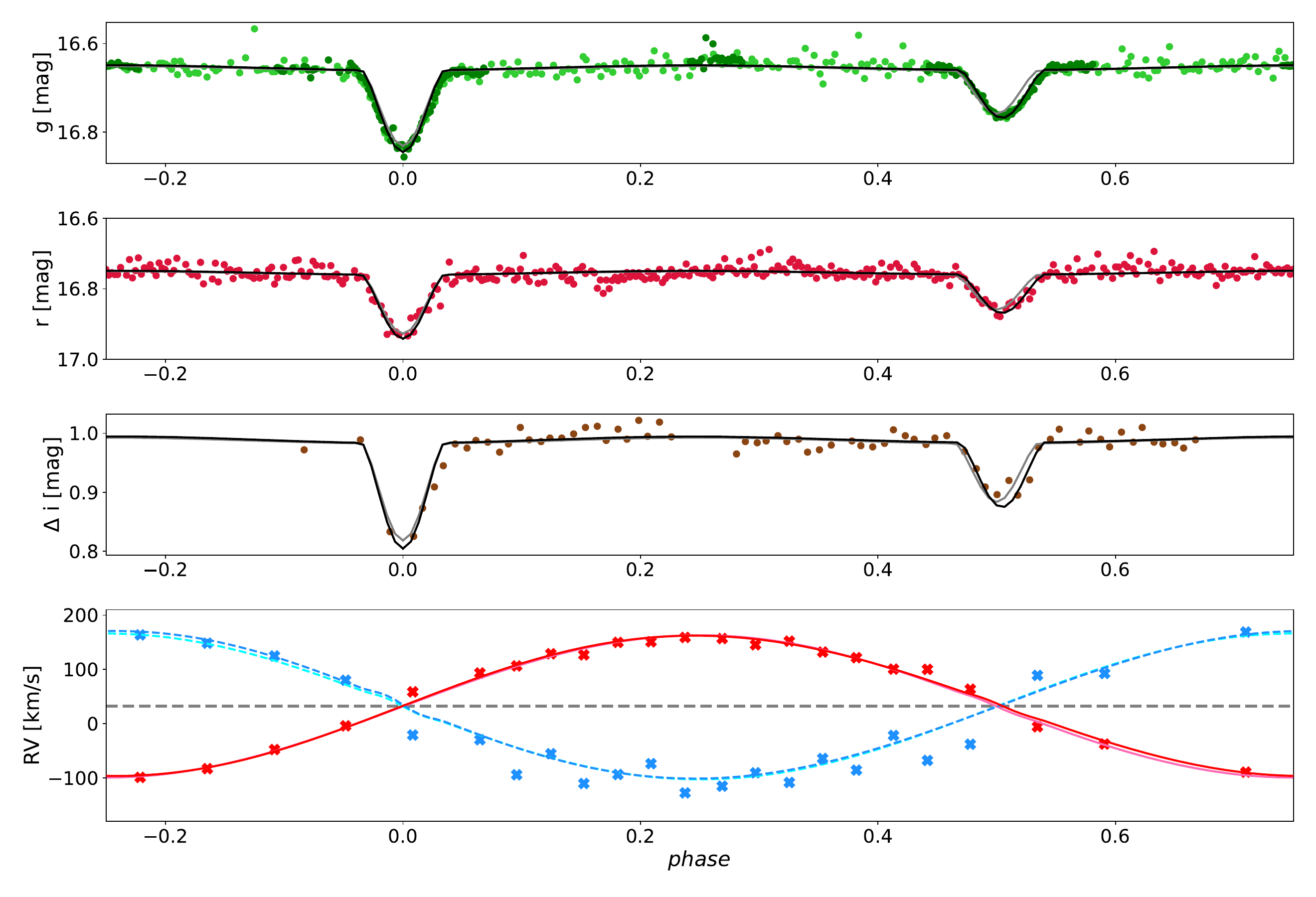}
\caption{The top three panels show our best fit eccentric PHOEBE model (black) compared to the observed ZTF g- (green) and r-band (red) and SARA i-band (brown) light curves. The gray line indicates our best model that does not account for eccentricity in the fit. 
The bottom panel shows the RVs of Star~1 (red crosses) and Star~2 (blue crosses) as measured assuming a surface ratio of 0.17. The formal fitting error bars for the assumed surface ratio (0.17) are smaller than the symbol sizes. The red and the pink lines indicated the best fit to the RV curve of Star~1 from the eccentric and circular model, respectively. For Star~2 the predicted RV curves from our best fit eccentric/circular model are indicated by the dashed blue/cyan lines. The gray dashed-dotted line indicates the system velocity.}
\label{fig:LC+RVs}
\end{figure*}

Given the large relative uncertainty on the \Gaia\ (E)DR3 distance, the fitting was not performed in absolute units, meaning that the fitting accounts for the depths and shapes of the eclipses but not the different brightnesses in the different filters (i.e. only the light curve morphology not the implied spectral energy distribution). As both stars are clearly hot and compact, extinction will have a minimal impact on the light curve morphology, nevertheless it was fixed in our modelling to the value determined in \se{sec:sed}.

Via Markov Chain Monte Carlo (MCMC) sampling of the model light and RV curves, we derived a best-fitting model. 
Initially, we fitted the RV curves of both stars, and let the radii and masses of both stars, \Teff and \logg of Star~2, the binary orbital inclination, and the system velocity vary freely. The derived \Teff and \logg of Star~2 as well as the surface ratio, $(R_2/R_1)^2$, were then fed back into the spectroscopic analysis.

Upon recognizing the sensitivity of the RV curve of Star~2 to the surface ratio adopted in the spectral fitting (see \se{sec:atmo}), we revised our modeling strategy. We stress that the surface ratio cannot be independently constrained from the light curves alone, but only the sum of their fractional radii $(R_1+R_2)/a$, where $a$ is the semi major axis of the binary orbit. Because of this subtle coupling between the surface ratio and the RV curve of Star~2, we consequently fixed the mass, temperature and surface gravity of Star~1 to the values determined in \se{sec:atmo}. This includes using the Kiel mass of $0.41 \pm 0.02$ \Msol (equivalent to a radius of $0.40\pm0.04$~\Rsol) determined by interpolating evolutionary tracks. Similarly, we fit the RV curves derived under the assumption of a surface ratio of $(R_2/R_1)^2 = 0.17$. This is well justified as the atmospheric parameters and the RV curve of Star~1 showed no significant variation across the individual spectroscopic fits employing different assumed surface ratios, and thus the model mass of Star~2 is principally dependent on the assumed Kiel mass of Star~1 and the inclination (constrained by the light curves).

Our best-fitting model is shown overlaid on the data in Fig.~\ref{fig:LC+RVs} with the parameters listed in Table~\ref{tab:para}.
In this final fit we derive a binary orbital inclination of $i=81.2_{-3.6}^{+0.6}$\,\textdegree, and RV semi-amplitudes of $K_1=129\pm9$\,km/s and $K_2=136\pm5$\,km/s for Star~1 and Star~2, respectively. For Star~2 we find a mass of $M_2=0.39\pm0.04$\,\Msol, an effective temperature of $75.0_{-22.0}^{+28.0}$~kK, and a surface gravity of $\log g = 5.60^{+0.30}_{-0.40}$ (equivalent to a radius of $0.16_{-0.07}^{+0.08}$~\Rsol). The resulting surface ratio (0.17) matches the one assumed in the spectroscopic fit (0.17). In addition, the predicted RV curve for Star~2 that is obtained from our final \textsc{phoebe2} model (dashed, blue line in the bottom panel of Fig.~\ref{fig:LC+RVs}), agrees well with the RV curve determined for Star~2 in the spectroscopic fit assuming $(R_2/R_1)^2=0.17$ (blue crosses in the bottom panel of Fig.~\ref{fig:LC+RVs}). Notably, also the system velocity obtained from the RV-curve fitting ($v_{\mathrm{sys}}=33.8\pm6.0$\,km/s) is in excellent agreement with the systemic velocity inferred from the nebular emission lines in \se{sec:v_sys} ($v_{\mathrm{sys}}=33.7\pm0.8$\,km/s). This consistency provides an important sanity check and supports the reliability of the RV curve (see also \citealt{Reindl+2020}).

Intriguingly, an eccentricity of $e =0.02$ is required to fit the light curve -- highlighted by the fact that the primary and secondary eclipses are not separated by exactly 0.5 in phase. This is particularly apparent in the secondary eclipse (during which Star~1 is obscured) as shown in Fig.~\ref{fig:LC+RVs}. The black line represent our best-fitting model, which includes the eccentricity of 0.02, whereas the light-gray curve shows the best-fitting model assuming a circular orbit (note that all parameters other than the eccentricity are within one sigma of the best-fitting model which allows for eccentricity). The circular model clearly predicts a secondary eclipse that occurs precisely at phase 0.5, appreciably earlier than observed. The impact of the eccentricity on the primary eclipse as well as the RV curves is less obvious. It is important to note that, given the long time coverage of the ZTF light curve and the excellent agreement with the RVs, it is highly unlikely that the implied eccentricity is a phasing issue due to an incorrect orbital period. 
R{\o}mer delay can also result in eclipses not being separated by 0.5 in phase, however the clearly similar masses (as implied by the RV curves) imply a negligible delay \citep{Barlow+2012} -- certainly not of the order of observed delay ($\approx$230s or 0.07 in phase).
Finally, we note that our best fitting model does show mild ellipsoidal modulation just as proposed by \cite{Chen+2025}.

\section{Spectral energy distribution} 
\label{sec:sed}

We performed a two-component fit to the observed spectral energy distribution (SED), employing the
$\chi ^2$ SED fitting routine described in \cite{Heber+2018} and \cite{Irrgang+2021}. 
We used photometry from \Gaia\/ (E)DR3 \citep{Gaia+2020, Gaia+2021}, 
Skymapper \citep{Onken+2019} and an the United Kingdom Infrared Telescope Hemisphere Survey (UHS, \citealt{Schneider+2025}). In the fit we assumed a surface ratio of 0.17, kept the atmospheric parameters fixed and let only the angular diameter, $\Theta$, and the color excess, $E(44-55)$,\footnote{\cite{Fitzpatrick+2019} employed $E(44-55)$, which is the monochromatic equivalent of the usual $E(B-V)$, using the wavelengths 4400\,\AA\ and 5500\,\AA, respectively. For high effective temperatures, such as the nucleus of Pa~13, $E(44-55)$ is identical to $E(B-V)$.} vary freely. Interstellar reddening was accounted for by using the reddening law of \cite{Fitzpatrick+2019}\footnote{We note that more recent reddening curves are available, e.g. \cite{Gordon+2023}.} with $R_V=3.1$. To compute the \Gaia\/ distance, a Gaussian distributed array of parallaxes was generated, with the distance being the median of this distribution (see \citealt{Irrgang+2021} for further details). The \Gaia\ parallax was corrected for the zeropoint bias using the Python code provided by \cite{Lindegren+2021}\footnote{\url{https://gitlab.com/icc-ub/public/gaiadr3_zeropoint}} and the parallax uncertainties are corrected by Eq.~16 in \cite{El-Badry+2021}.

We derive a reddening of \ebv$=0.405\pm0.005$\,mag, which is higher than what is suggested by the 2D dust map of \cite{Schlafly2011} who predicts \ebv$=0.339\pm0.011$\,mag. This indicates that the PN contributes to the reddening of the system.

From the angular diameter and parallax, the radii of the two stars can be calculated via \textit{R} = $\Theta$/2$\varpi$. By that we obtain $R_{\mathrm{Gaia}, 1}=0.45^{+0.32}_{-0.14}$\,$R_\odot$ and $R_{\mathrm{Gaia}, 2}=0.19^{+0.13}_{-0.06}$\,$R_\odot$, which agree well with the radii obtained from the light curve fitting.

\begin{figure}[t]
 \centering 
\resizebox{\hsize}{!}{\includegraphics{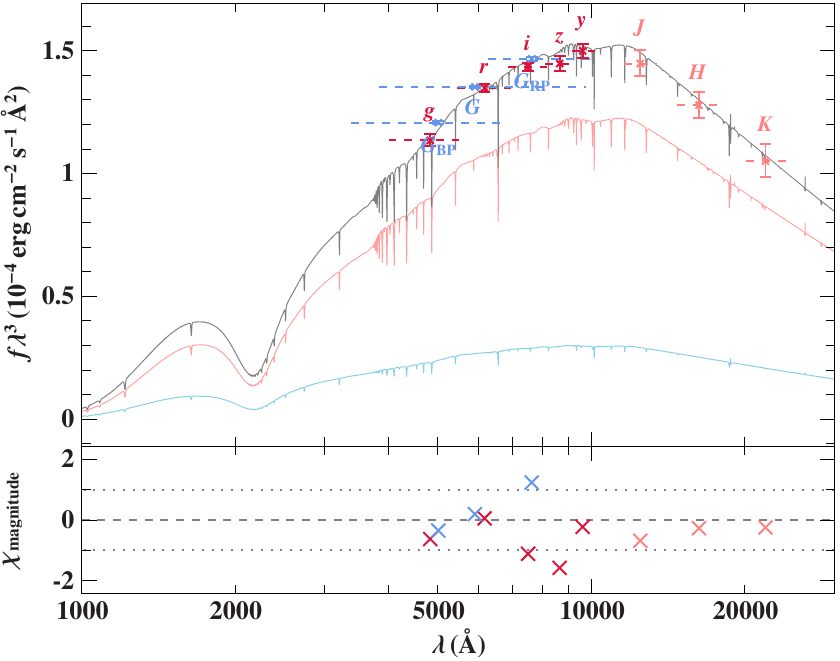}}
 \caption{Two-component spectral energy distribution fit. Top panel: Filter-averaged fluxes converted from observed magnitudes are shown in different colors (\Gaia: blue, Skymapper: red, UHS: pink). The light red and blue lines correspond to the flux contribution of Star~1 and Star~2, respectively, and the gray line is the combined best fit to the observation. The model fluxes are degraded to a spectral resolution of $6\,\AA$. To reduce the steep SED slope, the flux is multiplied by the wavelength cubed. Bottom panel: Uncertainty-scaled difference between synthetic and observed magnitudes.}
\label{fig:SED}
\end{figure}

\section{Kinematic analysis} 
\label{sec:kin}

\begin{figure}[t]
 \centering 
\resizebox{\hsize}{!}{\includegraphics{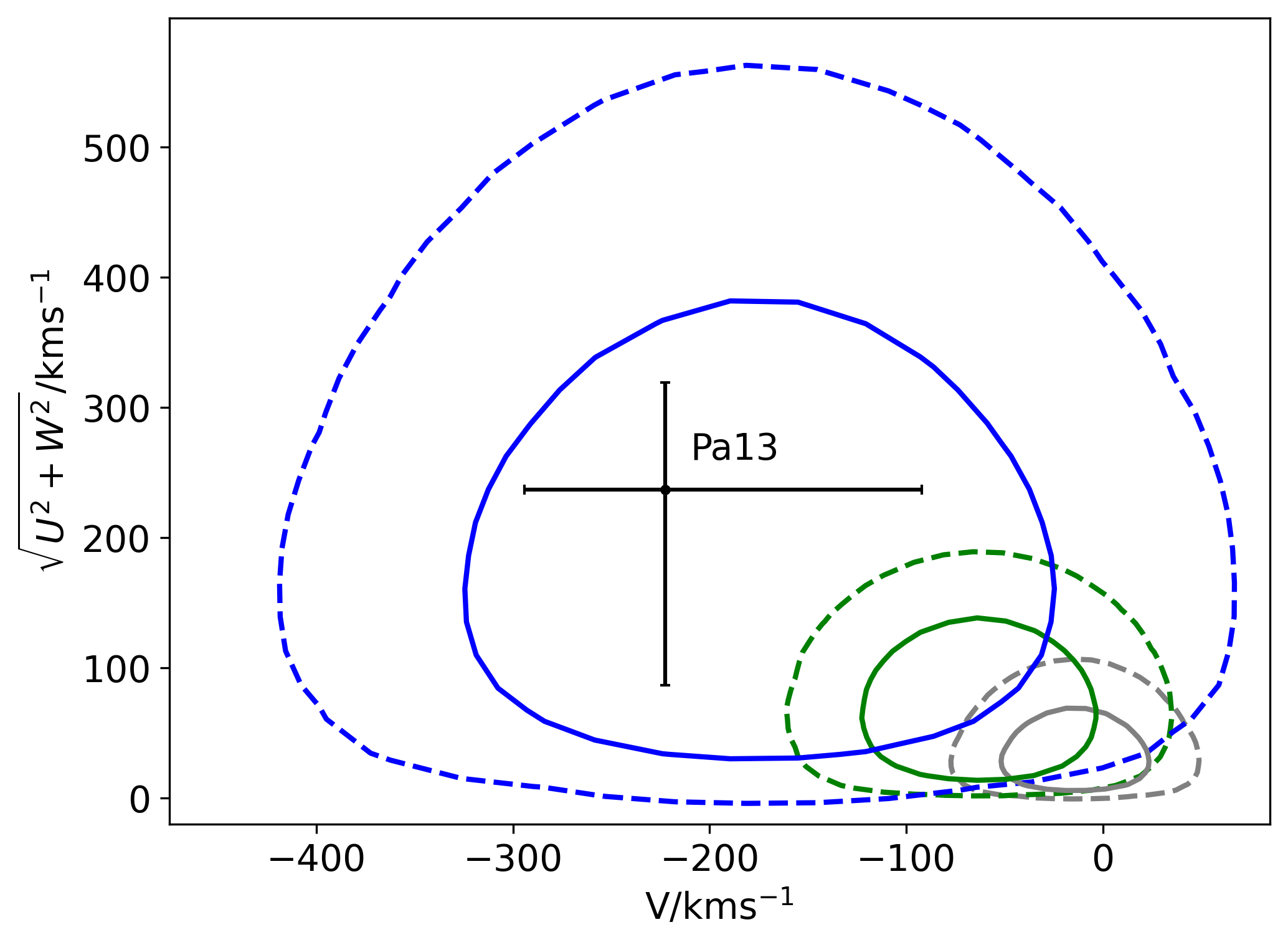}}
 \caption{Toomre diagram showing the location of Pa~13. The solid and dashed lines indicate the one- and two-sigma contours, respectively, of the U, V, and W velocity distributions of main-sequence stars from \cite{Kordopatis2011}. The gray, green, and blue lines indicate the velocity distribution of thin disk, thick disk, and Galactic halo stars, respectively.}
\label{fig:Toomre}
\end{figure}

Pa~13 is located at $z=-1.2^{+0.8}_{-0.4}$\,kpc below the Galactic plane, i.e. in a region where the thick disk dominates \citep{Kordopatis2011}. To further characterize the Galactic population membership of Pa~13, we calculated the space velocities ($U$ points away from the Galactic centre, $V$ points in the direction of Galactic rotation, and $W$ points toward Galactic north). This was done by using the system velocity derived in \se{sec:v_sys}, along with the proper motions and parallaxes from the \Gaia\ (E)DR3. From the Toomre diagram, which is shown in Fig~\ref{fig:Toomre}, it becomes clear that Pa~13 exhibits typical halo kinematics.

\section{Discussion} 
\label{sec:discussion}

\subsection{Eccentricity}

We also reported that Pa~13 shows clear evidence of an eccentric orbit ($e=0.02\pm0.01$). Similar levels of eccentricity have been found in several other post-CE systems containing more evolved WDs or hot subdwarf stars that lack a PN \citep{Kruckow+2021}. Among CSPNe, the only other system showing evidence of eccentricity is \object{NGC\,6026}. By modeling its light and RV curves, \cite{Hillwig+2010} found that the photometric variability is consistent with ellipsoidal modulation, where the CS nearly fills its Roche lobe and the companion is likely a hot WD. The light curves show unequal maxima, which -- in the case of ellipsoidal variability -- happens only at this amplitude for a non-circular orbit\footnote{Similarly appearing unequal maxima can occur in the case of Doppler boosting, but at amplitudes about ten times smaller than observed in NGC 6026.}. \cite{Hillwig+2010} showed that the CS light curve in NGC 6026 could be reproduced with an eccentricity in the range $e=0.01-0.03$.

The eccentricities observed in some post–CE binaries are thought to arise either from the pre-existing eccentricity \citep{Glanz+2021} or it may have been generated during the CE due to gravitational drag \citep{Szoelgyen+2022, RoepkeDeMarco2023}. One- and three-dimensional hydrodynamic simulations of CE evolution show that even if the pre-CE binary is initially circular, the resulting post-CE binary may have a non-zero eccentricity of order $e \approx 10^{-2}$ \citep{Sand+2020, Bronner+2024}, i.e. exactly what is observed for Pa~13 and \object{NGC\,6026}.

Interestingly, only recently the first two eccentric double-WD post-CE binaries have been reported by \citet{vanRoestel+2025}. Those authors find for \object{GALEX~J023834.9+093301} and \object{ZTF~J175812.85+764216.8} eccentricities of $e=(1.55\pm0.08)\times 10^{-3}$ and $e=(1.75\pm0.15)\times 10^{-3}$, respectively, which are about one order of magnitude lower than what is found for the two eccentric CSPNe. The cooling ages of the two eccentric double-WDs are of the order of $10^8$\,yrs, i.e. roughly four orders of magnitude larger than the post-CE ages of the eccentric CSPNe. The extreme rarity of eccentric double-WDs compared to younger, eccentric double-degenerate systems, may therefore reflect the progressive circularization of post-CE binaries over time.

\subsection{Formation of the system}

Our kinematic analysis has shown that Pa~13 most likely belongs to the Galactic halo. The age of the inner Galactic halo is estimated to be $10.9\pm0.4$\,Gyrs \citep{Kilic+2019}, which then should also correspond to the age of the Pa~13 binary system. 
With this information, we can exclude that both stars are low-mass ($\approx 0.33-0.5$\,\Msol) CO-core WDs. This is because the initial mass range to produce low-mass CO-core WDs is between approximately $1.8-3.0$\,\Msol \citep{PradaMoroni+2009}. The corresponding main-sequence life times of such stars at $Z=0.001$ is smaller than one Gyr according to BaSTI evolutionary models\footnote{\url{http://albione.oa-teramo.inaf.it/main_mod.php}}, i.e. about an order of magnitude smaller than the age of the inner Galactic halo.

Furthermore, we can rule out that the initially more massive star (from now on Star~A) evolved first to become a He-core WD and that the other, initially less massive Star~B recently turned into a low mass CO-core pre-WD. Let us assume Star~A with initial mass $M_{\mathrm{A}}$ first evolves into $\approx0.4$\,\Msol He-core WD. If we further presume that mass transfer was conservative, Star~B (with initial mass, $M_{\mathrm{B}}$), would have a post-mass-transfer mass of $M_{\mathrm{B}}'= M_{\mathrm{B}} + (M_{\mathrm{A}} -0.4$\,\Msol$)$. 
We require $M_{\mathrm{B}}'\gtrapprox 1.8$\,\Msol for the formation of a low mass CO-core WD, which again implies a main-sequence life time of less than one Gyr. In order to match the age of the inner Galactic halo, Star~A then would need to have had a main-sequence life time of $\approx 10$\,Gyrs, which would require $M_{\mathrm{A}} \approx 0.9$\,\Msol according to BaSTI evolutionary models. However, this would then imply Star~A was not be able to transfer enough mass to obtain $M_{\mathrm{B}}'\gtrapprox 1.8$\,\Msol.\\

Therefore, we conclude that the star that initiated the final CE episode must be a He-core pre-WD. By that, the PN must have been ejected by an RGB star. This provides hitherto the strongest evidence that the PNe phenomenon is not restricted to AGB stars, but that they can also be witnessed around post-RGB objects \citep{Jones+2023}.\\

Now, let us consider that both CSs of Pa~13 harbor a He-core. He-core WDs evolve from binaries in which both components had an initial mass of $\lessapprox 2.3$\,\Msol and filled their Roche lobes before helium ignition occurred in its degenerate core. Such systems may either form through "double spiral-in" -- also called double-core CE evolution -- or following two mass transfer episodes \citep{Nelemans+2001}.

\subsubsection{Double-core CE evolution}

For double-core CE evolution to occur, the components of the binary must possess almost equal initial masses. The reason for that is that the secondary must have left the main-sequence by the time the primary fills its Roche lobe, while the primary must do so before reaching the tip of the RGB \citep{Brown1995, Nelemans+2001, Dewi+2006, Justham+2011}. Initially, the system undergoes stable mass transfer, but when the secondary expands due to accretion, it ultimately overfills its Roche lobe. This prompts both stars to enter a CE phase at the same time and the successful ejection of the envelope leaves behind a compact binary consisting of the stripped cores of both components. Although double-core CE evolution is thought to be very rare given the highly constrained parameter space, recent observations have revealed a strong candidate for a progenitor system for this scenario (\object{BD$+20 5391$}, \citealt{Kurpas+2025}).

The mass ratio of Pa~13 is indeed close to unity, indicating that it may be a viable candidate for the double-core CE evolution scenario. Given the short lifetimes of PNe ($\approx 10^4$\,yrs) additional confirmation would be valuable for constraining the frequency of double-core evolution. We also note that this may not simply be a fortunate detection as an observational bias is expected among double-lined systems. Detecting double-lined, double-degenerate binaries generally requires both components to have comparable luminosities and post-RGB or post-AGB lifetimes, which in turn implies similar masses.

\subsubsection{Two mass transfer episodes with subsequent rejuvenation}

The more common case to form a close double He-core WD pair involves two mass transfer episodes. The initially more massive star evolves first, expands to fill its Roche lobe, and transfers mass to its companion. The physics of the first mass-transfer episode remains poorly constrained. \citet{Nelemans+2025} show that the commonly adopted prescriptions for stable mass transfer and CE evolution in binary population models do not simultaneously reproduce the observed double He-core WD population, indicating that the treatment of angular-momentum loss during these phases remains uncertain. In this scenario Star~A evolves first into a He-core WD and cools until Star~B starts to fill its Roche lobe. Mass loss then proceeds on a dynamical time scale, which is expected to trigger dynamically unstable mass transfer and the formation of a CE. This results in a drastic shrinkage of the orbit and the formation of a close, double He-core WD pair \citep{Nelemans+2001}.

Note that we cannot completely rule out the possibility of Star~A being low-mass CO WD or hybrid He-CO WD \citep{Zenati+2019} which formed after the first mass transfer episode. This would, however, imply a large difference between the main-sequence life times of both stars ($\Delta \tau_{\mathrm{MS}} \approx 10$\,Gyrs) given that they belong to the Galactic halo (see also above). This in turn would imply that Star~A then already had enough time to cool down to $\log L/$\Lsol$\lessapprox -4.0$ and $\Teff \lessapprox 6000$\,K \citep{PradaMoroni+2009}.

Regardless of whether Star~A is a He- or CO-core WD, a central prerequisite for the two-mass transfer episodes scenario is that the initially more massive Star~A needs to be effectively rejuvenated after the CE ejection, enabling it to evolve once again into a pre–WD configuration and to also stay in this configuration during the PN phase. Such a rejuvenation process of the WD that formed after the first mass transfer episode was also proposed for \object{Hen\,2-428}, the only other double-lined, double-eclipsing, double-pre-WD system inside a PN. The formation of the system was explained by a first stable mass transfer episode which resulted in a 0.35\,\Msol He-core WD and a 2.5\,\Msol main-sequence star \citep{Reindl+2020}. After about 500\,Myr, the latter evolved up the early AGB, a CE was formed, and later ejected. That implies that the He-core WD had enough time to cool down to $\approx 0.3$\,\Lsol and $\Teff \approx 25$\,kK \citep{Hall+2013}, yet it is also found in a pre-WD configuration with \Teff$=40$\,kK and $L=665$\,\Lsol. 

In the case of Pa\,13, even if Star~B evolved away from the main sequence soon after the first mass transfer event, it would take the red giant between 100\,Myr and $\approx 1$ Gyr to form a He-core of $\approx 0.4$\,\Msol \citep{MillerBertolami2016}. During this time the core of Star A would have already cooled and faded down significantly. Consequently, we expect the rejuvenated WD to harbor a much cooler core than those of freshly born post-RGB stars. These rejuvenated objects with cold cores are known to be brighter, and evolve faster, than "normal" post-RGB stars (e.g., \citealt{2011MNRAS.415.1396M, Althaus+2013, Istrate+2014}). The higher \Teff of Star 2 in Pa~13 makes it the most likely candidate for the rejuvenated star in the system (i.e. Star~A). If this is the case, then Star 1 in Pa~13 would be the last to evolve (i.e. Star~B) and the one with normal post-RGB core.

\subsection{Post–CE evolutionary state and envelope masses}

A reliable characterization of the evolutionary status of both stars based solely on their surface parameters (\Teff, \logg, and $\log L$) is not possible due to the current lack of understanding of the CE event. Thermal equilibrium models like those computed by \cite{Hall+2013} can predict the post-RGB heating/contracting timescales (as well as the corresponding tracks) robustly for stars with a young (hot) He-core\footnote{By robustly we mean specifically that, as long as the object is in the so-called "thermal equilibrium", and as long as the core has not cooled down from its RGB temperature, both the tracks and the heating rates will not deviate from those computed by \cite{Hall+2013}.}, but the starting point of the post-CE evolution ($T_{\rm{eff}_0}$, $g_0$, $R_0$) in the Hertzsprung–Russell or Kiel diagram depends on the mass of the H envelope that survives the CE. Current models of the CE event do not provide us with this quantity. Noteworthy, in the case of Pa\,13 the presence of a PN around the system might help us to shed light on the immediate post-CE conditions.

Assuming a distance of $6.9$\,kpc, a PN diameter of 29\,arcsec (as listed in the HASH database) and a typical expansion velocity of 20\,km/s \citep{Schoenberner+2018}, a kinematical age of $24$\,kyrs can be calculated for the PN. For the typical inferred mass of both stars ($\approx 0.4$\,\Msol), the 0.403\,\Msol track from \cite{Hall+2013} predicts a rather constant heating rate of $d T_{\rm eff}/dt\simeq 0.65$ K/yr (during the evolution at constant luminosity). This, together with the derived post-CE age from the PN would imply that the effective temperatures of the stars in Pa~13 immediately after the CE ($T_{\rm{eff}_0}$) have been $\approx 15.6$\,kK lower than today. Notably, if this is the case, it would mean that immediately after the CE Star~1 in Pa~13 had a radius of $R_0\simeq 0.8\ R_\odot$, which is exactly the size of its Roche lobe at its current separation \citep[$a=2.1\ R_\odot$, $q\simeq 1$,][]{1983ApJ...268..368E}.

Therefore, it is possible to speculate that when the PN was ejected during the CE approximately 24~kyr ago, both stars were still filling their Roche lobes in an over-contact configuration (with $T_{\rm{eff}_0}\approx 35$\,kK). The system subsequently became detached as both components contracted, with Star~2 contracting faster due to its possibly rejuvenated nature. Within this picture, detached systems such as Pa~13 would be more evolved analogs of over-contact systems like \object{Hen\,2-428}.\\

Interestingly, the derived immediate post-CE effective temperatures, $T_{\rm{eff}_0}$, can be used to estimate the envelope masses immediately after the CE. This is because, for a given core mass, the starting point of the post-CE evolution in the Kiel/Hertzsprung–Russell diagram ($T_{\rm{eff}_0}$, $g_0$, $R_0$) of a thermal equilibrium remnant is solely determined by its envelope mass. As envelope masses are not provided by \cite{Hall+2013} for their post-RGB models we computed our own $0.4$\Msol post-RGB model using the {\tt LPCODE} code \citep{MillerBertolami2016, MillerBertolami+2022}. Our $0.4$\Msol model reproduces very closely the post-RGB evolution computed by \cite{Hall+2013}. These models predict that Star~1 and Star~2 have been left with H-rich envelopes of masses $M_{\rm H}^1\approx 1.72\times 10^{-3}$\Msol and $M_{\rm H}^2\approx 1.25\times 10^{-3}$\Msol, respectively, immediately after the CE event\footnote{These masses are very likely underestimated since the post-RGB models were calculated for $Z=0.02$ enabling a direct comparison with the models shared by P.~D. Halls. At lower metallicity (which is expected for Pa~13), larger envelope masses are to be expected, yet they will still be significantly smaller than the corresponding peel-off mass.}.

It is worthwhile to mention that these envelopes of masses are well below the critical "peel-off" masses\footnote{The peel-off mass is defined as the maximum mass of a contracting thermal equilibrium remnant. A thermal equilibrium remnant with mass greater than this is a red giant and expands as its core mass grows \citep{Hall+2013}.} for these stars ($M_{\rm H}\simeq 2.68\times 10^{-3}$\Msol for a star departing the RGB at $5000$K, see \citealt{Hall+2013} for a detailed discussion). This indicates that envelope masses after the CE are very likely smaller than the peel-off mass. Consequently, characterizing double degenerate CSPNe offers an interesting avenue to constrain the mass ejection process during the CE.

Of course, these estimations should not be taken at face value for two main reasons. Firstly, within current uncertainties in the mass, heating timescales can vary by a factor of two. For example \cite{Hall+2013} predicts heating rates of $d T_{\rm eff}/dt\simeq 1.2$ K/yr and 0.32\,K/yr for their 0.427\,\Msol and 0.378\,\Msol post-RGB models, respectively. This will affect the estimation the immediate post-CE effective temperatures. Moreover, if one of the two CSs is a rejuvenated object the use of post-RGB models with hot cores will lead to an underestimation of the heating rate. Rejuvenated post-RGB-like structures -- such as He-core WDs undergoing a H-shell flash -- have a burning shell on top of a more compact He-core. These objects are known to be brighter and harbor thinner H-rich envelopes, which in turn leads to much faster heating rates (e.g., \citep{2011MNRAS.415.1396M,Althaus+2013,Istrate+2014, Althaus+2025}).

Regardless of these uncertainties, the previous discussion highlights the fact that young post-CE systems that are still surrounded by a PN might allow us to infer critical information of the CE event itself. In particular, in the case of Pa\,13 at least one of the two stars has to be a young post-RGB star (i.e. one that that is properly described by the models computed by \citealt{Hall+2013}) that was left after the CE with an envelope mass significantly less massive than the peel-off mass. This result offers a key constraint for hydrodynamical models of the CE ejection.

\subsection{Fate of the system}

\begin{figure}[t]
 \centering 
\resizebox{\hsize}{!}{\includegraphics{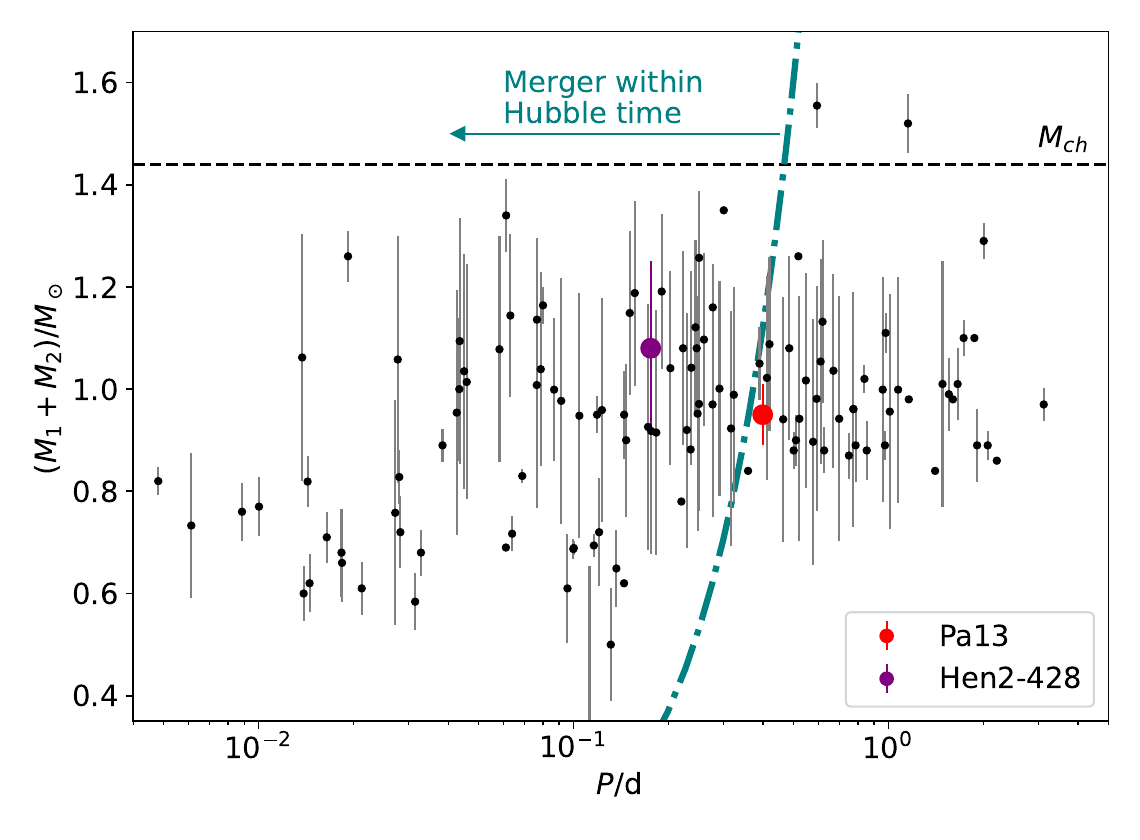}}
 \caption{Total mass plotted against logarithmic period of DD systems. Pa\,13 is shown in red and Hen\,2-428 in purple. The black dots are close double WD systems compiled by \cite{Munday+2024}. The teal, dashed-dotted line indicates which systems will merge within a Hubble time, and the black, dashed line indicates the Chandrasekhar mass limit.}
\label{fig:hubble}
\end{figure}

Last but not least, we would like to comment on the fate of the system. Since the eccentricity of the system is small, we calculate the Peters' time-scale, $\tau_{\mathrm{P}}$ (\citealt{Peters1964}, see also \citealt{Zwick+2020}) to estimate the time the orbit will decay due to the radiation of gravitational waves via:
$$\tau_{\mathrm{P}} = \frac{5\,c^5(1+q)^2 a_0}{256\,G^3M^3 q f(e_0)},$$
where $c$ is the speed of light, $G$ the gravitational constant, $q = M_2/M_1 \leq 1$ the mass ratio of the system, $M$ the total system mass, $a_0$ the initial Keplerian semi major axis, and $f(e_0)$ is the correction of the merger time due to the initial eccentricity of the orbit, $e_0$ and can be calculated via:
$$f(e) = \left(1 + \frac{73}{24} e^2 + \frac{37}{96} e^4\right) (1-e^2)^{-7/2}.$$
By that we estimate a merger time of $22.8^{+12.0}_{\phantom{0}-7.8}$\,Gyrs. This implies that the system will not merge within a Hubble time, which is also illustrated in Fig~\ref{fig:hubble}.

Yet when the system eventually merges, a possible outcome is the formation of a highly magnetic ($\approx 200$–$500$ kG) He-sdO. Strikingly, all known magnetic He-sdOs have inferred masses of about 0.8\,\Msol \citep{Dorsch+2022, Dorsch+2024, Pelisoli+2022}, which is close to the maximum mass expected from the merger of two He-core WDs and about the total mass of the CSs of Pa~13. About one percent of all He-sdOs exhibit magnetic fields stronger than 100\,kG, and it has been suggested that this small fraction may be linked to the requirement that the progenitor He-WDs have coeval masses \citep{Pakmor+2024}.

\section{Conclusion} 
\label{sec:conclusion}

In this work we reported that Pa~13 is the second double eclipsing and double-lined, pre-WD system inside a PN, but the first detached such system.
Furthermore, we found that Pa~13 provides hitherto the strongest evidence that the PN phenomenon is not restricted to post-AGB stars, but can also be observed around post-RGB stars. The high quality light curves allowed us to derive an orbital eccentricity of $0.02\pm0.01$ -- making Pa~13 the second post-CE binary CS with a measured eccentricity.

Since the mass ratio of Pa~13 is close to unity the system may have formed through double-core CE evolution. Alternatively, the system may have formed following two mass transfer episodes. In the latter case systems like Pa~13 and Hen~2-428 would indicate that there must exist an efficient CE-induced rejuvenation mechanism capable of reheating the cool WD in the binary, possibly an extreme analog to the observed heating and inflation observed in main sequence star companions of post-CE CSPNe. 

Assuming a kinematical age of 24\,kyr, we infer that immediately after the CE ejection the effective temperatures of both components of Pa~13 were $\approx 15.6$\,kK lower than today and that Star~1 in Pa~13 must still have filled its Roche lobe. This suggests that Pa 13 represents a more evolved, detached descendant of over-contact double-degenerate systems such as \object{Hen\,2-428}.

The inferred immediate post-CE effective temperatures, $T_{\rm{eff}_0}$, also allowed us to place direct constraints on the hydrogen-envelope masses of the two components in Pa~13, since $T_{\rm{eff}_0}$ of post-RGB models of a given mass is mainly dictated by the residual envelope mass. We estimated post-CE envelope masses which are about a factor two smaller than the typical peel-off mass for a post-RGB remnant. This demonstrates that double-degenerate systems still embedded within PNe offer particularly powerful constraints of the system's immediate post-CE configuration (e.g. $T_{\rm{eff}_0}$, $g_0$, $R_0$) and a direct means of assessing envelope masses. Expanding observational efforts to identify and characterize additional double-degenerate systems inside PNe, is thus strongly encouraged.

For Pa~13, higher SNR spectroscopic observations, higher cadence and SNR light curves, and a framework capable of simultaneously fitting spectra, light curves, and RV curves would help to better constrain the system parameters in the future. Follow-up imaging and spectroscopy of the PN are also encouraged to resolve its morphology and to determine its expansion velocity and chemistry. In addition, evolutionary models capable of reproducing the formation of this system would be highly desirable.

\begin{acknowledgements}
We thank the Anonymous Referee for their careful review and valuable suggestions. We thank Stephan Geier for helpful discussion.
N.R. is supported by the Deutsche Forschungsgemeinschaft (DFG) through grant RE3915/2-1. T.H. acknowledges support from the National Science Foundation: this material is based upon work supported by the National Science Foundation under Award No. AST2107768. Any opinions, findings and conclusions or recommendations expressed in this material are those of the authors and do not necessarily reflect the views of the National Science
Foundation. D.J. acknowledges support from the Agencia Estatal de Investigaci\'on del Ministerio de Ciencia, Innovaci\'on y Universidades (MCIU/AEI) under grant ``Nebulosas planetarias como clave para comprender la evoluci\'on de estrellas binarias'' and the European Regional Development Fund (ERDF) with reference PID-2022-136653NA-I00 (DOI:10.13039/501100011033). DJ also acknowledges support from the Agencia Estatal de Investigaci\'on del Ministerio de Ciencia, Innovaci\'on y Universidades (MCIU/AEI) under grant ``Revolucionando el conocimiento de la evoluci\'on de estrellas poco masivas'' and the European Union NextGenerationEU/PRTR with reference CNS2023-143910 (DOI:10.13039/501100011033).
M3B acknowledges support from CONICET and Agencia I+D+i through grants PIP-2971 and PICT 2020-03316, as well as from the CONICET-DAAD 2022 bilateral cooperation grant number 80726.
Based on observations made with the GTC telescope, in the Spanish Observatorio del Roque de los Muchachos of the Instituto de Astrofísica de Canarias, under Director’s Discretionary Time. 
This paper uses data taken with the MODS spectrographs built with funding from NSF grant AST-9987045 and the NSF Telescope System Instrumentation Program (TSIP), with additional funds from the Ohio Board of Regents and the Ohio State University Office of Research. 
Based on observations collected at the European Organisation for Astronomical Research in the Southern Hemisphere under ESO programmes 0109.D-0237(A) and 0111.D-2143(A).
IRAF is distributed by the National Optical Astronomy Observatory, which is operated by the Association of Universities for Research in Astronomy (AURA) under a cooperative agreement with the National Science Foundation.
This work has made use of BaSTI web tools.
This research has made use of the HASH PN database at \url{http://hashpn.space}.
This research made use of the SIMBAD database, operated at CDS, Strasbourg, France; the VizieR catalogue access tool, CDS, Strasbourg, France. This research has made use of the services of the ESO Science Archive Facility.
This work has made use of data from the European Space Agency (ESA) mission \Gaia\ (https://www.cosmos.esa.int/gaia), processed by the \Gaia\ Data Processing and Analysis Consortium (DPAC, https://www.cosmos.esa.int/web/gaia/dpac/consortium). Funding for the DPAC has been provided by national institutions, in particular the institutions participating in the \Gaia\ Multilateral Agreement.
Based on observations obtained with the Samuel Oschin 48-inch Telescope
at the Palomar Observatory as part of the Zwicky Transient Facility
project. ZTF is supported by the National Science Foundation under Grant
No. AST-1440341 and a collaboration including Caltech, IPAC, the
Weizmann Institute for Science, the Oskar Klein Center at Stockholm
University, the University of Maryland, the University of Washington,
Deutsches Elektronen-Synchrotron and Humboldt University, Los Alamos
National Laboratories, the TANGO Consortium of Taiwan, the University of
Wisconsin at Milwaukee, and Lawrence Berkeley National Laboratories.
Operations are conducted by COO, IPAC, and UW.

\end{acknowledgements}

\bibliographystyle{aa}
\bibliography{BB}

\begin{onecolumn}
\begin{appendix}

\section{Figures}

\begin{figure*}[ht]
\includegraphics[width=\textwidth]{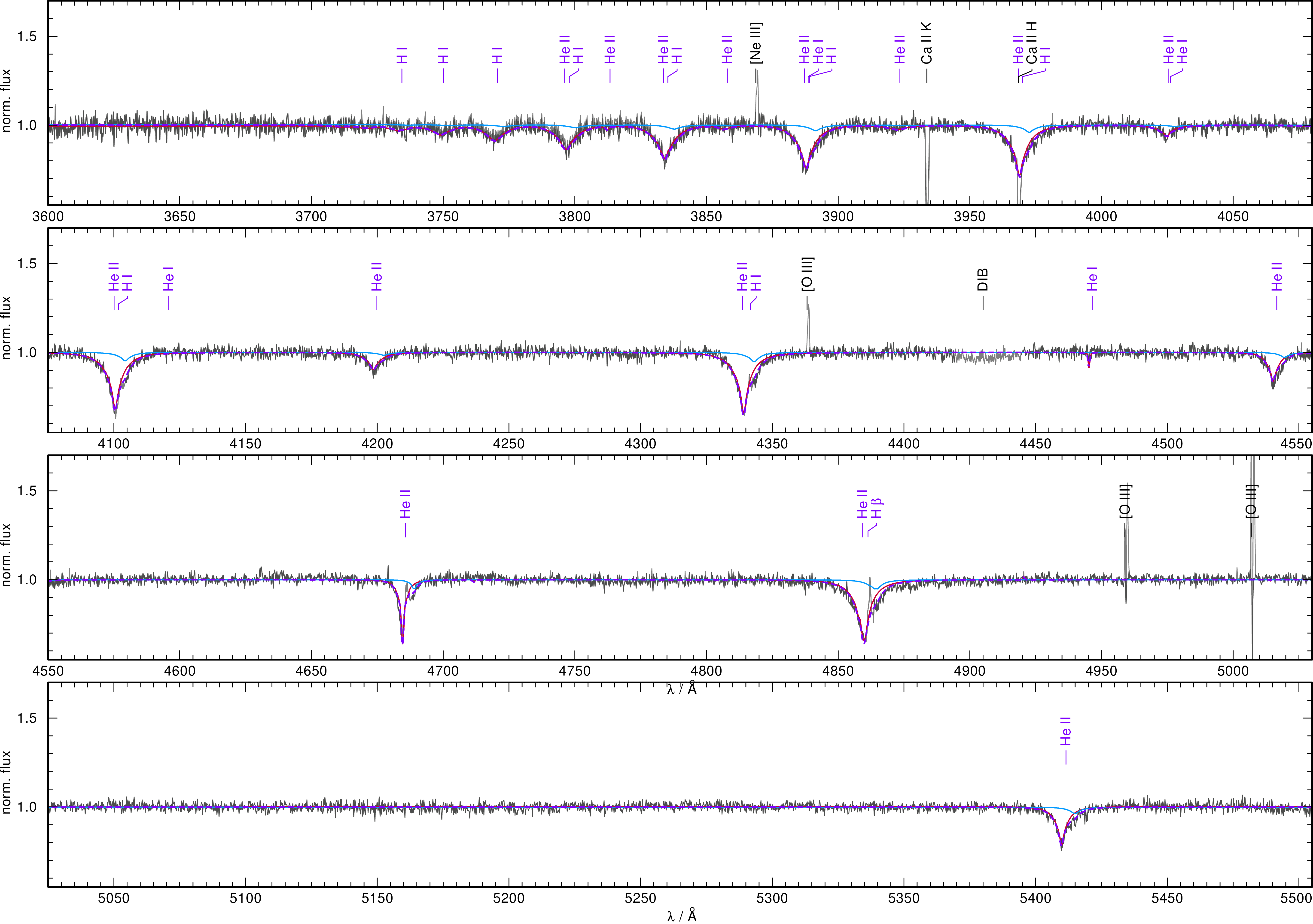}
\caption{UVB X-Shooter spectrum taken at phase 0.84 (gray) compared to our best-fitting TMAP models assuming a surface ratio of 0.17. Light gray regions indicate the location of a weak residual nebular and interstellar lines, which have been excluded from the fit. The red and blue lines represent the contribution of Star~1 and Star~2, respectively, and the dashed, purple line shows the combined fit combined fit.}
\label{fig:ubv}
\end{figure*}

\end{appendix}

\end{onecolumn}

\end{document}